\documentclass[pra,notitlepage,twocolumn,amsmath,amssymb,aps,superscriptaddress,10pt]{revtex4-2}
\usepackage{amsmath, amssymb, amsfonts, bbm, times}
\usepackage[dvipsnames]{xcolor}
\usepackage[T1]{fontenc}
\usepackage{braket}
\usepackage[normalem]{ulem}
\usepackage[colorlinks=true,bookmarks=false,linkcolor=RoyalBlue,urlcolor=RoyalBlue,citecolor=RoyalBlue,breaklinks]{hyperref}
\usepackage{graphicx}
\usepackage{bm}
\usepackage{physics}

\begin{document}

\title{Continuous-time noise mitigation in analogue quantum simulation}
\author{Gabriele Bressanini}
\affiliation{Blackett Laboratory, Imperial College London, London SW7 2AZ, United Kingdom}
\author{Yue Ma}
\affiliation{Blackett Laboratory, Imperial College London, London SW7 2AZ, United Kingdom}
\author{Hyukjoon Kwon}
\affiliation{Korea Institute for Advanced Study, Seoul 02455, South Korea}
\author{M.S. Kim}
\affiliation{Blackett Laboratory, Imperial College London, London SW7 2AZ, United Kingdom}
\affiliation{Korea Institute for Advanced Study, Seoul 02455, South Korea}

\begin{abstract}
Analogue quantum simulators offer a promising route to explore quantum many-body dynamics beyond classical reach in the near term.
However, their vulnerability to noise
limits the accuracy of simulations.
Here, we establish a new framework for mitigating noise in analogue quantum simulation, operating in a time-continuous manner.
To our knowledge, this is the first protocol that is fully analogue and that achieves exact noise cancellation.
Our method requires a small number of ancillary qubits,
whose interaction with the system$-$combined with classical post-processing of joint measurement data$-$is tailored to cancel the effect of noise.
Furthermore, the protocol is Hamiltonian-independent, robust to realistic ancilla noise, and avoids any discretization, preserving the continuous-time nature of the system's dynamics. 
This work opens a new direction for achieving high-fidelity analogue quantum simulation
in the presence of noise.
\end{abstract}

\maketitle

\section*{Introduction}

One of the most promising applications of quantum technology$-$originally envisioned by Feynman \cite{feynman}$-$is the efficient simulation of the dynamics of highly correlated many-body quantum systems, which are central to problems in physics, chemistry, and materials science \cite{daley2022practical, SimulationReview, cirac2012goals, PRXQuantum.2.017003}.
A typical quantum simulation task involves preparing an initial state, evolving it under a target dynamics, and measuring the expectation value of a relevant observable.
The target dynamics is, in most cases, the unitary evolution generated by a Hamiltonian, though certain applications require the simulation of non-unitary open system dynamics governed by a Lindblad equation \cite{Plenio_Lindblad}. 
Digital, gate-based quantum computers,
combined with quantum error correction (QEC), will eventually achieve fault-tolerance and enable reliable simulation of arbitrary Hamiltonian dynamics. Nevertheless, despite promising recent results \cite{PhysRevX.11.041058,PRXQuantum.5.030326,ai2024quantum,google2023suppressing,bluvstein2024logical}, large-scale QEC remains challenging for current devices.
On the other hand, analogue quantum simulators offer a realistic path to beyond-classical results in the near term, even without achieving fault-tolerant universal quantum computing.
They can already leverage current hardware capabilities, provided that the target Hamiltonian can be mapped onto the simulator's Hamiltonian, either exactly or with controllable error.
Moreover, while digital quantum simulators require discretization of time evolution, typically via Trotterization, which can introduce systematic errors and increase compilation overhead \cite{lloyd1996universal, PhysRevX.11.011020}, analogue simulators naturally realize continuous-time dynamics. This may provide substantial near-term advantages in simulating larger systems over longer times, with experimental proposals spanning a variety of physical platforms such as cold atoms in optical lattices, trapped ions, and superconducting qubits \cite{esslinger2010fermi,houck2012chip,PhysRevLett.92.207901,blatt2012quantum,PhysRevLett.93.250405,zohar2015quantum,PhysRevLett.109.125302,PhysRevLett.110.055302,ebadi2021quantum,RevModPhys.93.025001}.

A major challenge in analogue simulators is their inherent lack of error correction, which limits the size and timescale of simulated systems, as well as the achievable accuracy of measurement outcomes.
\begin{figure}
    \centering
    \includegraphics[width=\linewidth]{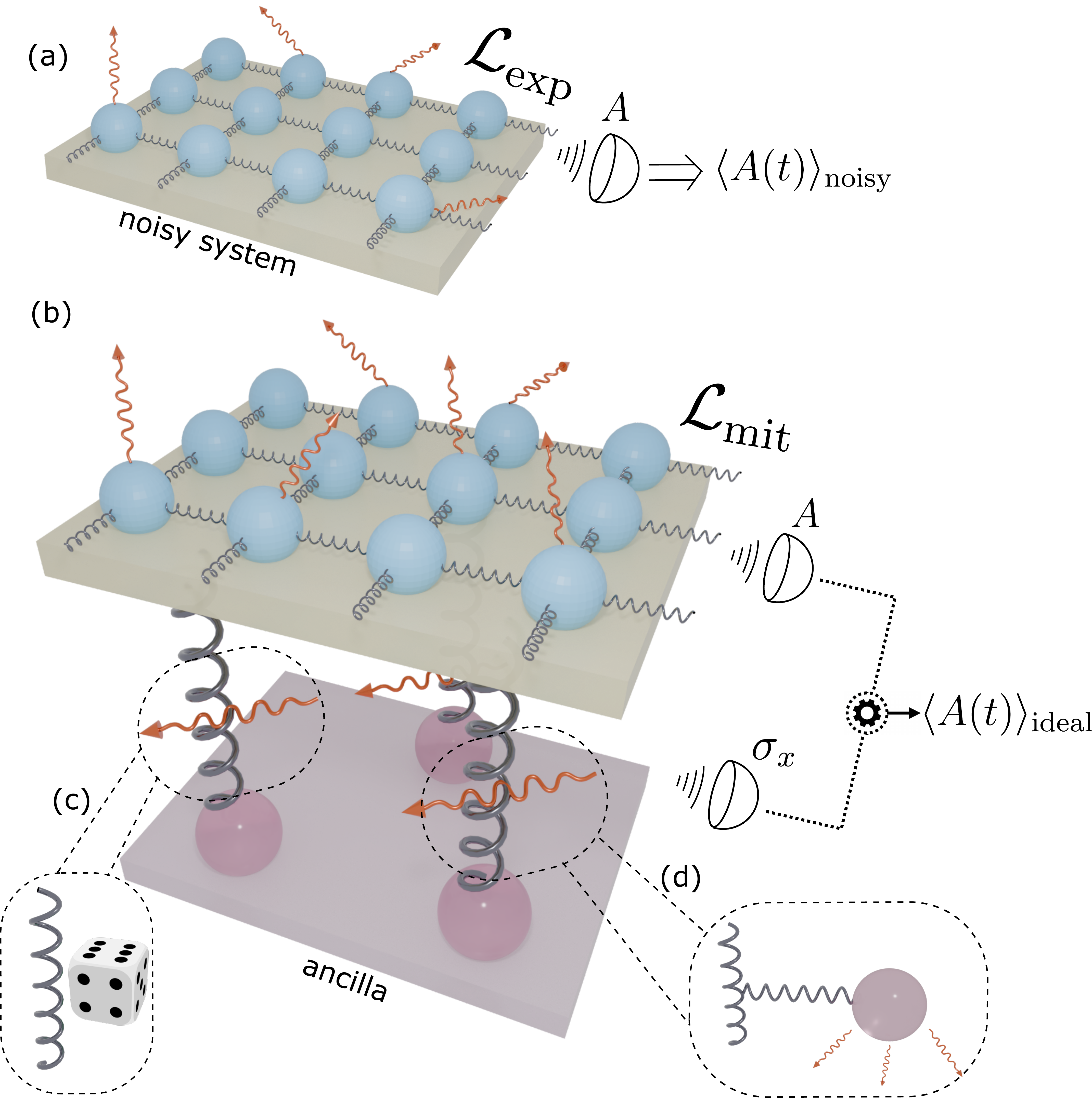}
    \caption{Schematics of the noise mitigation protocol and its implementation. (a) A quantum system evolves under noisy experimental dynamics described by the Lindbladian $\mathcal{L}_\textrm{exp}$ in Eq.~\eqref{experimental_lindblad_equation} for time $t$, and then an observable $A$ is measured. 
    (b) In the mitigation scheme, the system is coupled to one or more ancillary qubits, each prepared in the state $\ket{+}$ and evolves under the joint dissipation, described by $\mathcal{L}_\textrm{mit}$ in Eq.~\eqref{lindbladian_mitigation}.
    After evolving the joint system to time $t$, joint measurements of $A$ on the system and $\sigma_x$ on the ancillae are combined through classical post-processing to yield an unbiased estimation of the noise-free expectation value $\langle A(t)\rangle_\textrm{ideal}$.
    The joint dissipation between system and ancilla can be implemented either using (c) a stochastic Hamiltonian or (d) coupling to an additional fast-decaying ancilla.
    }
    \label{fig:schematics}
\end{figure}
As an alternative, a feasible near-term approach in analogue quantum simulators would be quantum error mitigation (QEM)~\cite{QEMReview}, which focuses on reducing noise-induced bias in expectation values—often the key quantities of interest in present-day experiments—through repeated noisy circuit sampling and classical post-processing, making it suited to noisy intermediate-scale quantum (NISQ) devices.
However, most QEM protocols found in the literature are explicitly tailored to discrete, gate‑based digital simulators, where noise is assumed to act either immediately before or immediately after each gate;
a prominent example is given by probabilistic error cancellation (PEC) \cite{PEC_Temme}. 
To apply these techniques to analogue quantum simulation, one must approximate the continuous-time evolution using a sequence of discrete time steps. However, such an approach reintroduces the discretization errors that analogue simulation is designed to avoid, which leads to residual errors that grow with simulation time.
While significant effort has been devoted to understanding how errors accumulate over time in observable expectation values measured on analogue quantum simulators \cite{cai2023stochastic,PRXQuantum.1.020308,PhysRevA.109.052425,trivedi2024quantum}, no QEM protocol that is both natively analogue and truly time-continuous has yet been proposed. A stochastic QEM protocol inspired by PEC~\cite{JinzhaoMitigationPaper} requires interleaving the simulator’s time evolution with discrete recovery operations at predetermined time sequences, which is
more naturally suited to hybrid analog-digital architectures.
A recently proposed trajectory-based error mitigation scheme~\cite{trajectoryQEM} utilizes repeated measurement and reinitialization of an ancilla qubit for continuously monitoring quantum jumps, which is also a demanding task in analogue quantum simulators.

In this work, we introduce
a fully time-continuous error cancellation protocol designed specifically for analogue quantum simulators. Our method is native to the analogue setting, as it preserves the system’s continuous-time evolution by operating entirely within the framework of Markovian open system dynamics, described by the Lindblad master equation, enabling exact noise cancellation.
The protocol consists of coupling the system to a small number of ancillary qubits, with a single qubit being sufficient in principle, and allowing the system–ancilla state to evolve under the Lindblad master equation involving additional joint jump operators. The joint state is then measured, and a classical post-processing step is applied to recover the ideal noiseless expectation value of an arbitrary system observable (see Fig.~\ref{fig:schematics}). Remarkably, the proposed protocol is also robust against typical and experimentally relevant forms of noise affecting the ancilla qubit.

The proposed protocol offers two key advantages over previously introduced schemes~\cite{JinzhaoMitigationPaper, trajectoryQEM}, as it neither relies on gate-based control with discretized time steps, nor requires intermediate measurements on the system or the ancilla during the evolution. In addition, the protocol is entirely independent of the specific Hamiltonian being simulated, ensuring universality across diverse quantum simulation tasks and physical platforms.
We discuss the physical implementation of our protocol and show that, for a broad class of commonly employed noise models, it can be readily realized via ensemble averaging over a stochastic Hamiltonian.
We demonstrate the protocol's effectiveness through three representative: the Heisenberg model, the Loschmidt echo following a quantum quench, and the Floquet dynamics of a periodically driven spin chain. Our approach opens a new pathway toward high-fidelity analogue quantum simulations resilient to decoherence.

\section*{Results}

\subsection*{Framework}
Let us consider a finite-dimensional system initialized in state $\rho(0)$ at $t=0$, whose evolution is governed by the following Lindblad master equation
\begin{equation}
    \label{experimental_lindblad_equation}
    \frac{d}{dt}\rho=-i[H,\rho]+\sum_k \mathcal{D}[L_k](\rho) \equiv \mathcal{H}(\rho) + \mathcal{D}(\rho) \equiv \mathcal{L}_\textrm{exp}(\rho) \, ,
\end{equation}
where $\mathcal{D}[L](\cdot)=L\cdot L^\dag-\frac{1}{2}\lbrace L^\dag L,\cdot\rbrace$ denotes the dissipator superoperator.
The Lindbladian $\mathcal{L}_\textrm{exp}$ models a typical experimental dynamics with Markovian noise, where $H$ is a target Hamiltonian of interest and $L_k$ are jump operators corresponding to noise that result from the interaction between the system and the environment. In general, the Hamiltonian and the noise operators may have an explicit time dependence, but in what follows, we will assume them to be time-independent for simplicity. The expectation value of a quantum observable $A$ at time $t$ reads $\expval{A(t)}_\textrm{noisy} = \Tr[A e^{\mathcal{L}_\textrm{exp}t}(\rho(0))]$, which is inherently affected by the noise.  
The central question we address--and answer affirmatively--is whether it is possible to eliminate the effects of noise and recover the ideal, noiseless expectation value $\expval{A(t)}_\textrm{ideal} = \Tr[A e^{-iHt}\rho(0)e^{iHt}]$.  

The effects of noise and the Hamiltonian do not, in general, commute with each other, i.e., $e^{\mathcal{H}t+\mathcal{D}t} \neq  e^{\mathcal{D}t}e^{\mathcal{H}t}$. As a result, noise can propagate non-trivially throughout the dynamics, rendering the error-mitigation task particularly challenging.
Previous approaches~\cite{JinzhaoMitigationPaper, trajectoryQEM} to tackling this problem have relied on discretising either the system’s evolution, resulting in approximate protocols, or the recovery operations, yielding schemes that are better suited to hybrid analogue–digital architectures.

\subsection*{Noise mitigation protocol}
\label{sec:protocol}
We present our continuous-time noise mitigation protocol, which is independent of both the (possibly time-dependent) system Hamiltonian and the observable of interest $A$.
The protocol requires only a single-qubit ancilla, regardless of the system’s dimension.
We aim to perfectly cancel the noisy contributions to the experimental dynamics in Eq.~\eqref{experimental_lindblad_equation} by introducing an additional \emph{negative} dissipator, i.e. $-\mathcal{D}(\rho)$, leaving us with the ideal unitary dynamics generated by the Schr\"odinger equation $\dot{\rho} =-i[H,\rho]$.
However, the negative sign in front of this superoperator renders the latter non-physical and therefore not directly implementable.

We overcome this problem by introducing a qubit ancilla that interacts with the system through joint dissipation, inspired by techniques that encode the system's dynamics in the off-diagonal blocks of the system–ancilla state.~\cite{Aspuru_Hamiltonian_Sim, GenericMapLindbladToNonLindblad, Non-Markovian_sim}. Upon projecting the entire system-ancilla state after their joint dissipation onto the subspace spanned by the off-diagonal elements in the ancilla's computational basis, we obtain an operator that evolves according to the noiseless Schr\"{o}dinger equation of the system.

For an explicit description of the protocol, let $W$ be the system-ancilla joint density matrix, and consider the following Lindblad master equation
\begin{equation}
    \label{recovery_master_equation}
    \frac{d}{dt}W = \mathcal{L}_{\textrm{exp}}(W)+ \mathcal{L}_\textrm{mit}(W) \, .
\end{equation}
Here, $\mathcal{L}_\textrm{exp}$ acts trivially on the ancilla qubit and $\mathcal{L}_\textrm{mit}$ denotes the mitigation Lindbladian 
\begin{equation}
    \label{lindbladian_mitigation}
    \mathcal{L}_\textrm{mit} = \sum_k \mathcal{D}[L_k\otimes\sigma_z] + \mathcal{D}[\sqrt{S}\otimes{\sigma_z}] + \mathcal{D}[\sqrt{S}\otimes{\mathbb{I}}] \, ,
\end{equation}
where $\mathbb{I}$ and $\sigma_z = \ketbra{0}-\ketbra{1}$ are the identity and the Pauli $Z$ operators, respectively, and $S = a\mathbb{I}-\sum_k L^\dag_k L_k \geq 0$, with $a = \lambda_{\textrm{max}}(\sum_k L_k^\dag L_k)$ being the largest eigenvalue of $\sum_k L_k^\dag L_k$.
We now project Eq.~\eqref{recovery_master_equation} onto the computational basis of the ancilla qubit, defining $W_{ij}\equiv \bra{i}W\ket{j}$, where $i,j \in\lbrace 0,1 \rbrace$.
These four operators act on the system Hilbert space, and satisfy the following equations (see Methods for a detailed derivation)
\begin{align}
\label{schrodinger_with_exp_decay}
\frac{d}{dt}W_{ij} &= -i[H,W_{ij}] - 2aW_{ij} \, , \\
\label{diagonal_equation}
\frac{d}{dt}W_{ii} &= -i[H,W_{ii}] + 2\sum_k \mathcal{D}[L_k](W_{ii}) + 2\mathcal{D}[\sqrt{S}](W_{ii}),
\end{align}
for $i\neq j$.
We observe that the joint dissipators introduced by $\mathcal{L}_\textrm{mit}$ have the effect of removing the original system noise terms from the off-diagonal blocks, Eq.~\eqref{schrodinger_with_exp_decay}, while doubling the noise contributions in the diagonal ones, Eq.~\eqref{diagonal_equation}. 
In particular, Eq.~\eqref{schrodinger_with_exp_decay} coincides with the noiseless Schr\"odinger equation, up to an additional exponential decay term given by $-2aW_{ij}$.
Consequently, if we initialize the system and ancilla in the state $W(0) = \rho(0) \otimes \ketbra{+}$, with $\ket{+}=(\ket{0}+\ket{1})/\sqrt{2}$, the solution to Eq.~\eqref{schrodinger_with_exp_decay} becomes
\begin{equation}
    W_{01}(t) = \frac{e^{-2at}}{2}e^{-iHt}\rho(0)e^{iHt} = W_{10}(t) .
\end{equation}
Hence, $W_{ij}(t)$ corresponds to the ideal system state evolved under the noise-free unitary dynamics, up to an exponential damping prefactor. Finally, as shown in Methods,
we can express the ideal expectation value of $A$ as 
\begin{equation}
\begin{split}
    \label{mitigated_exp_val}
    \expval{A(t)}_\textrm{ideal} =e^{2at} \Tr[(A\otimes\sigma_x)W(t)] \, ,
\end{split}
\end{equation}
where $\sigma_x = \ketbra{0}{1} + \ketbra{1}{0} $ denotes the Pauli $X$ operator.
We can estimate the decay rate ${a}$ by setting $A = \mathbb{I}$, since
$
1 = \expval{\mathbb{I}(t)}_\textrm{ideal} = e^{2at} \Tr[(\mathbb{I}\otimes\sigma_x)W(t)].
$
In other words, evaluating $\Tr[(\mathbb{I}\otimes\sigma_x)W(t)]$ directly gives information about the decay exponent, which we then use to express a mitigated expectation value as
\begin{equation}
    \label{alternative_exp_value_formula}
    \expval{A(t)}_\textrm{ideal} = \frac{\Tr[(A\otimes\sigma_x)W(t)]}{\Tr[(\mathbb{I}\otimes\sigma_x)W(t)]} \, .
\end{equation}

If the noise operators have explicit time dependence, the previous derivation still holds, with the modification that $a$ is promoted to a function of time, defined as $a(t)=\lambda_\textrm{max}(\sum_kL_k^\dag (t) L_k (t))$.
In this case, the ideal expectation value can be written as $\expval{A(t)}_\textrm{ideal} = e^{2\int_0^t a(s)ds} \Tr[(A\otimes\sigma_x)W(t)]$.

As a special case, if the noise operators satisfy $\sum_k L_k^\dag L_k \propto\mathbb{I}$, setting $a = \lambda_\textrm{max}(\sum_k L_k^\dag L_k)$ causes $S$ to vanish, reducing Eq.~\eqref{lindbladian_mitigation} to the simpler form
\begin{equation}
    \label{simplified_mitigation_lindbladian}
    \mathcal{L}_\textrm{mit} = \sum_k \mathcal{D}[L_k\otimes\sigma_z] \, .
\end{equation}
In this case, the mitigation Lindbladian can be implemented even without explicitly computing $a$.
A typical example of such a situation is when each noise operator is a 
single qubit Pauli operator.
This also greatly enhances experimental implementability, as we only need to realize each pair of joint dissipations between a single system qubit and the ancilla qubit.

We can summarize the protocol for obtaining the noiseless expectation value of an arbitrary observable $A$ at time $t$ $-$given that the system evolves according to the noisy dynamics in Eq.~\eqref{experimental_lindblad_equation}$-$ as follows: 
\begin{enumerate}
    \item Prepare the ancilla in the $\ket{+}$ state,  
    \item Implement the joint system Lindblad evolution described by Eq.~\eqref{recovery_master_equation} for the same time $t$,  
    \item Evaluate the expectation value of the observable $A \otimes \sigma_x$ for the joint system,  
    \item Multiply the result by the prefactor $e^{2at}$.  
\end{enumerate}

While, in principle, our protocol only requires a single ancilla qubit, such an implementation may be experimentally challenging, as the ancilla must interact with all system qubits.
This requirement can be difficult to satisfy in platforms with limited connectivity, and may demand the engineering of non-trivial many-body interactions.
To address these constraints, in the Supplementary Information, we explore several equivalent alternative realizations of our scheme, including protocols employing multiple ancilla qubits, up to the limiting case where each system qubit is paired and interacts with its own ancilla, as well as extensions involving qutrit ancillae and/or modified dissipative couplings.

\subsection*{Sampling overhead}
As is typical with error mitigation techniques, removing bias comes at the cost of increased variance~\cite{QEMReview} in the estimator of the expectation value.
From Eq.~\eqref{mitigated_exp_val}, it is clear that the protocol's sampling overhead, defined as the scaling factor between the variance of the mitigated estimator and that of the unmitigated (noisy) estimator at a fixed number of measurement shots $n$, originates from the exponential pre-factor $e^{2at}$.

The noisy estimator of the expectation value of $A$ at time $t$ simply reads  $\hat{A}_\textrm{noisy} (t) =n^{-1}\sum_{i=1}^n A_i(t)$, 
where $A_i(t)$ denotes the measurement outcome from the $i$th shot and the system evolving  according to Eq.~\eqref{experimental_lindblad_equation} up to time $t$.
On the other hand, the mitigated unbiased estimator is $\hat{A}_{\textrm{mit}}(t) = n^{-1} e^{2at} \sum_{i=1}^n {A_i (t)\sigma_{x,i}(t)}$, 
where $A_i(t)\sigma_{x,i}(t)$ denotes the $i$th measurement outcome of the observable $A\otimes\sigma_x$ on $W(t)$.
Therefore, assuming that the per-shot variances are comparable in both the mitigated and unmitigated scenarios, the sampling overhead is approximately $e^{4at}$.
Since $a = \lambda_\textrm{max}(\sum_k L_k^\dag L_k)$, it follows that the sampling overhead scales exponentially with both the total noise rate and the evolution time, an expected feature of all QEM protocols.
Applying Hoeffding's inequality, one finds the number of samples required to ensure the mitigated estimator is $\epsilon-$close to its expectation value $\expval{A(t)}_\textrm{ideal}$ with probability no less than $1-\delta$ scales as $ n \propto e^{4at} {\log(2/\delta)}/{\epsilon^2}$ .

\subsection*{Protocol robustness against ancilla noise}
\label{subsec:ancillanoise}
In any realistic implementation of the protocol, the ancilla qubit would also be subject to noise. Therefore, it is crucial to evaluate the effect of this additional noise source on our protocol. 
Remarkably, we show that a simple rescaling of the measurement outcomes can correct for most ancilla noise models. To illustrate this, let us consider a single noise operator $M$ acting on the qubit ancilla.
The Lindblad equation governing the evolution of the system-ancilla joint state is then given by
\begin{equation}
    \label{recovery_equation_with_ancilla_noise}
    \frac{d}{dt}W=\mathcal{L}_\textrm{exp}(W) + \mathcal{L}_\textrm{mit}(W) + \mathcal{L}_\textrm{anc}(W) \, ,
\end{equation}
with $\mathcal{L}_\textrm{anc}(W)= \Gamma \mathcal{D}[\mathbb{I}\otimes M](W)$, where $\Gamma$ is the ancilla noise rate. Consequently, the differential equation that governs the evolution of $W_{ij}$ described in Eq.~\eqref{schrodinger_with_exp_decay} now contains the additional term
\begin{equation}
    \label{ancilla_contribution}
    \Gamma \bra{i}\mathcal{D}[\mathbb{I}\otimes M](W)\ket{j} \, .
\end{equation}
Although the operator $M$ generally depends on the specific physical implementation of the qubit, we can nonetheless compute this additional contribution for widely used and realistic noise models, such as dephasing ($\sigma_z$ or $\ketbra{0}$ and $\ketbra{1}$), amplitude damping ($\sigma_-$), thermal excitation ($\sigma_+$) or Pauli Lindblad noise channels $(\sigma_x,\sigma_y,\sigma_z)$. 
These noise operators essentially encompass all the commonly used models for describing single-qubit noise in practical settings.


By substituting all these noise operators into Eq.~\eqref{ancilla_contribution}, we find that the contribution takes the form $-\nu\Gamma W_{ij}$, where $\nu$ is a non-negative constant determined by the specific jump operator.
In particular, the contribution vanishes ($\nu=0$) for $\sigma_x$, while we obtain $\nu=2$ for $\sigma_z$ and $\sigma_y$, and finally we have $\nu=1/2$ for the remaining operators $\sigma_\pm, \ketbra{0},$ and $\ketbra{1}$.

As a result, even with ancilla noise, Eq.~\eqref{schrodinger_with_exp_decay} maintains the form of a Schr\"odinger equation with exponential damping, the only modification being an increased decay constant.
Crucially, the recovery protocol remains unchanged: $\mathcal{L}_\textrm{mit}$ is unaffected by the ancilla noise, which is instead accounted for by means of a simple classical post-processing, namely an appropriate redefinition of the exponential pre-factor $e^{2at}\mapsto e^{2\tilde{a}t}$ in Eq.~\eqref{mitigated_exp_val}, which only affects the sampling overhead of the scheme.
We stress that only the exponential pre-factor is modified, while the parameter $a$ appearing in the operator $S$ within $\mathcal{L}_\textrm{mit}$ remains unchanged.

If the ancilla qubit is affected by multiple noise processes corresponding to different noise operators from the ones described above, the mitigation scheme remains valid, provided that the effective decay rate $\tilde{a}$ is adjusted to account for the combined contributions of all ancilla noise operators.
Once again, we can estimate the effective decay rate $\tilde{a}$ via 
$
e^{-2\tilde{a}t}  =  \Tr[(\mathbb{I}\otimes\sigma_x)W(t)]
$, hence Eq.~\eqref{alternative_exp_value_formula} still holds.

In the Supplementary Information we show that our mitigation scheme can also handle certain forms of correlated noise between the system and the ancilla, such as $\mathcal{D}[P\otimes \sigma_\pm]$, where $P$ is a generic Pauli string acting on the system qubits.

\subsection*{Protocol robustness against imperfect noise characterization}
\label{sec:robustness}
An important and practical aspect to investigate is the protocol’s robustness to imperfect noise characterization. Specifically, we assume that the Lindblad noise operator is known, but the associated rate $\gamma$ is only approximately estimated, with $\tilde{\gamma} = \gamma + \epsilon$ denoting the estimated rate and $\epsilon$ a small error. 
We also assume perfect implementation of any desired Lindbladian. In the following, we separately examine how imperfect knowledge of the noise, whether on the system or the ancilla, affects the accuracy of our protocol.
We begin by analysing the impact of imperfect knowledge of the system's noise. 
Let us assume that the noise is modelled by a single Lindblad noise operator $\sqrt{\gamma}L$, for which we have learned an approximation $\sqrt{\tilde{\gamma}}L$. 
Straightforward calculations yield the evolution equation for $W_{ij}$
\begin{equation}
    \label{residual_noise_equation}
    \frac{d}{dt}W_{ij} =-i[H,W_{ij}] -\epsilon D[L](W_{ij}) - 2aW_{ij} \, ,
\end{equation}
showing that, although perfect noise cancellation is not achieved, an accurate estimate of $\gamma$ ensures that most of the noise is effectively mitigated.
Eq.~\eqref{residual_noise_equation} implies that, in this scenario, our protocol yields estimates of  expectation values corresponding to the system's dynamics being governed by a master equation of the same form as Eq.~\eqref{experimental_lindblad_equation}.
Interestingly, when $\epsilon<0$, the master equation reduces to the initial Lindblad equation, but with an effective noise rate of $\gamma-\tilde{\gamma}$. On the other hand, if $\epsilon > 0$, the resulting equation corresponds to a non-physical, pseudo-Lindblad equation, which describes non-Markovian dynamics.

We now turn our attention to the case where the ancilla noise rate is only approximately known. As before, the true noise operator is $\sqrt{\gamma}M$, while our learned approximation is $\sqrt{{\gamma}+\epsilon}M$.
Assuming that $M$ is a correctable noise operator, $\mathcal{L}_\textrm{mit}$ remains unchanged, and the only adjustment is a modification of the exponential pre-factor, resulting in an overall multiplicative pre-factor that scales like  $e^{\nu\epsilon t}$, for some noise-dependent constant $\nu$.

\subsection*{Physical implementation}
We first consider the case where the system's noise jump operators $L_k$ are Hermitian and unitary, such as Pauli operators, and the mitigation protocol is defined by the jump operators in Eq.~\eqref{simplified_mitigation_lindbladian}. This scenario is particularly relevant for current quantum hardware, where Pauli noise constitutes a dominant source of errors.
In this setting, the effective dynamics generated by each jump operator $L_k \otimes \sigma_z$ can be realized by taking the ensemble average over a stochastic Hamiltonian~\cite{Budini00non-markovian, chenu2017quantum},
\begin{equation}
    H_{\rm int}(t) = \sum_{k} \eta_k(t) \left( L_k  \otimes \sigma_z \right),
\end{equation}
where the coefficients $\eta_k (t)$	are independent white noise processes. 
This stochastic Hamiltonian approach has already been demonstrated experimentally in trapped-ion systems~\cite{Maier19environment} and superconducting qubits~\cite{Li25observation}, highlighting the near-term feasibility of testing the scheme in this simplified regime.
Beyond the Pauli noise case, more general joint evolutions of the system and ancilla can be realized by various different dissipation engineering techniques~\cite{Verstraete09quantum, chenu2017quantum, Harrington22engineered}. A standard way to implement each jump operator of the mitigation Lindbladian ${\cal L}_{\rm mit}$ in Eq.~\eqref{lindbladian_mitigation} is to couple the entire system with another fast-decaying ancilla system with the interaction Hamiltonian~\cite{Verstraete09quantum}.
\begin{equation}
    H_{\rm int} \propto L^\dagger_{\rm mit} \otimes \sigma^{(a)}_- + L_{\rm mit} \otimes \sigma^{(a)}_+,
\end{equation}
where $L_{\rm mit}$ is the target jump operator for mitigation, i.e., $L_k\otimes\sigma_z$, $\sqrt{S}\otimes{\sigma_z}$, and $\sqrt{S}\otimes{\mathbb{I}}$ in Eq.~\eqref{lindbladian_mitigation}, and $\sigma^{(a)}_\pm$ are the lowering and raising operators of the fast-decaying ancilla state with dissipator ${\cal D}[\sigma^{(a)}_-]$. While this requires multi-qubit interactions between system and ancillary states, such an operation has been recently realized in trapped-ion~\cite{van2024experimental} and quantum dot~\cite{zapusek2022nonunitary} systems.

\subsection*{Examples}
\subsubsection{Heisenberg model simulation}
As a first example, we consider a 2D anisotropic Heisenberg model on a $2\times 2$ square lattice, with Hamiltonian
\begin{equation}
    H = \sum_{\expval{ij}} \left[ J_x \sigma_x^{(i)} \sigma_x^{(j)} + J_y \sigma_y^{(i)}  \sigma_y^{(j)} + J_z \sigma_z^{(i)} \sigma_z^{(j)} \right] - h \sum_{i=1}^4 \sigma_y^{(i)},
\end{equation}
where $\expval{ij}$ denotes nearest-neighbour pairs, $h$ is the strength of the transverse field, and the superscripts indicate the spins on which the operators act. We further take $J_x = J(1+\gamma)$, $J_y = J(1-\gamma)$, and $J_z = J$ with the anisotropy parameter $\gamma$ and the strength of the pairwise interaction $J$. Such a model has been extensively studied to investigate quantum magnetism and criticality \cite{PhysRevB.44.446,torelli2018calculating,PhysRevB.61.14601,PhysRevLett.52.1579}.

We assume that each qubit in the analogue simulator$-$including both system and ancilla$-$is subject to dephasing and relaxation noise, modelled by the jump operators $\sqrt{\gamma_z}\sigma_z$ and $\sqrt{\gamma_-}\sigma_-$, respectively.
We are interested in the expectation value of the total magnetization, i.e., $\mathcal{M} =\sum_{i=1}^4\sigma_z^{(i)}$.
To construct the jump operators for recovery,
we compute the decay constant $a$ and the operator $\sqrt{S}$. Specifically, we find $a = \lambda_\textrm{max}(\sum_{i=1}^4 ( \gamma_z\sigma_z^{(i)\dag}\sigma_z^{(i)} + \gamma_-\sigma_-^{(i)\dag}\sigma_-^{(i)})) 
= 4(\gamma_z + \gamma_-)$ and, using that $S$ is diagonal in the computational basis, we may write $\sqrt{S}$ in the following compact form
\begin{equation}
    \sqrt{S}=\sum_{{x\in\lbrace0,1\rbrace^4}} \sqrt{4-h(x)}\ketbra{x} \, ,
\end{equation}
where $h(x)$ denotes the Hamming weight of the bit-string $x$, i.e. the number of its non-zero digits.
The four system qubits and the ancilla qubit are initialized in the states $\ket{0}$ and $\ket{+}$, respectively. 

\begin{figure}
    \centering
     \includegraphics[width=\linewidth]{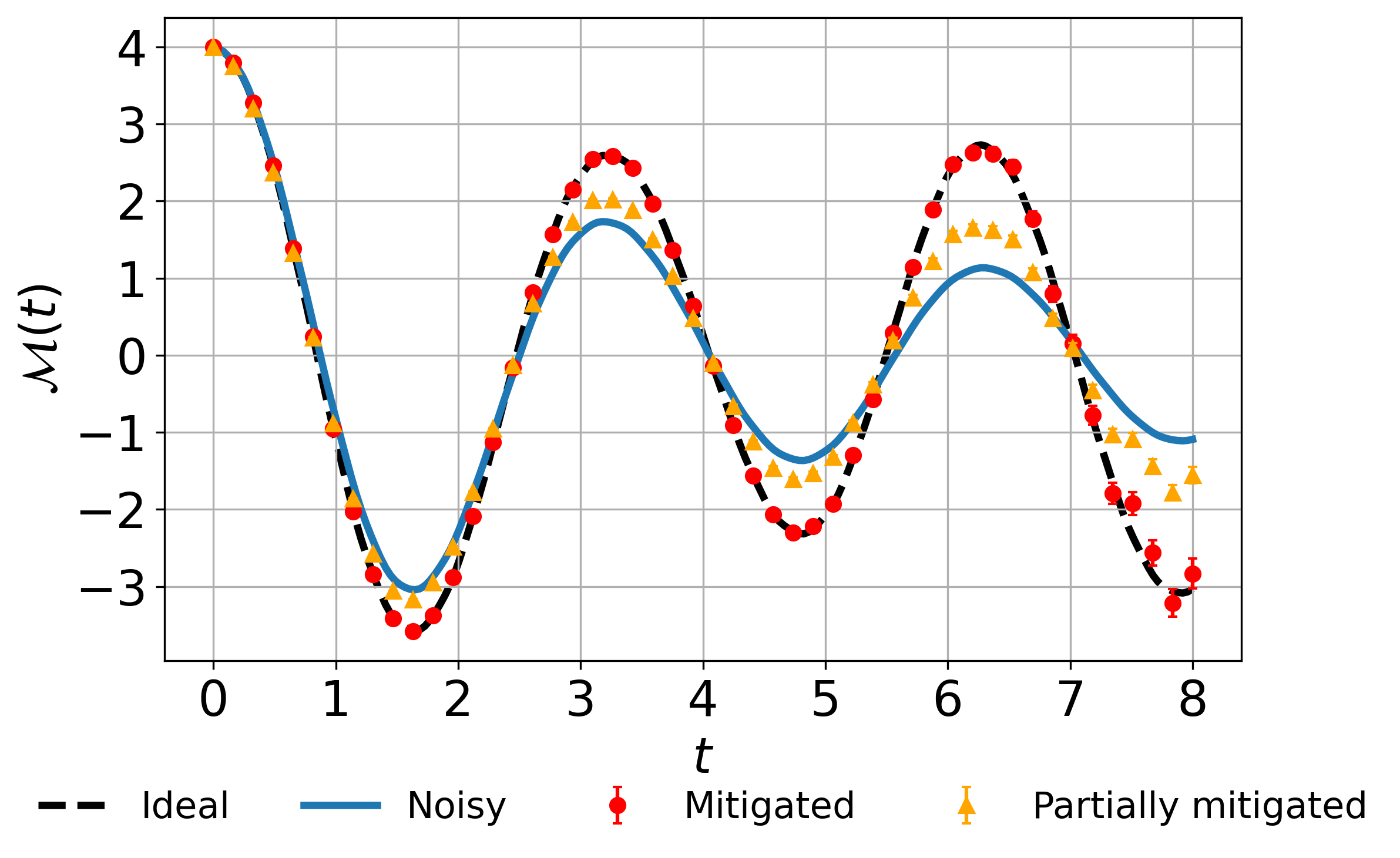}
    \caption{Expectation value of the total magnetization in the 2D anisotropic Heisenberg model with transverse field and $N=4$ spins. The mitigated expectation values and their standard deviations (represented by the error bars) are obtained by averaging $10^6$ samples. Partial mitigation corresponds to the correction of system noise only, ignoring ancilla noise.}
    \label{fig:Heisenberg}
\end{figure}

We let the joint state evolve under Eq.~\eqref{recovery_equation_with_ancilla_noise}, after which we perform a joint measurement of the total magnetization on the system qubits and $\sigma_x$ on the ancillary qubit.
Using Eq.~\eqref{mitigated_exp_val}, the noise-free expectation value of $\mathcal{M}$ can be expressed as 
\begin{equation}
    \label{Heisenberg_mitigated_magnetization}
    \expval{\mathcal{M}(t)}_\textrm{ideal}= e^{(2(4\gamma_z+4\gamma_-)+(2\gamma_z + \frac{1}{2}\gamma_-))t} \Tr[(\mathcal{M}\otimes\sigma_x)W(t)] \, ,
\end{equation}
where $2\gamma_z+ \frac{1}{2} \gamma_-$ in the exponential pre-factor corrects the ancilla noise, as previously discussed. 
In Fig.~\ref{fig:Heisenberg} we plot the noisy, ideal, and mitigated expectation values of the total magnetization, for $\gamma = 0.2$, $J = 2$, $h = 0.1$, and $\gamma_z = \gamma_- = 0.03$.
The mitigated expectation values at each time step are estimated using $10^6$ samples. 
Notice that when ancilla noise is not accounted for in the mitigation scheme, the resulting multiplicative error grows exponentially with time, leading to a substantial discrepancy from the ideal expectation value.

\subsubsection{Loschmidt echo}
Next, we apply our protocol to retrieve characteristic features of quench dynamics that would otherwise be obscured by noise.
A quantum quench refers to a non-equilibrium scenario where a system is initially prepared in the ground state $\ket{\psi_0}$ of an initial Hamiltonian $H_0$, which is then suddenly changed$-$typically through a rapid variation of a tunable parameter$-$to a new Hamiltonian $H$, resulting in the following dynamics $\ket{\psi(t)} = e^{-iHt}\ket{\psi_0}$. 
Following a quantum quench, the system may exhibit critical behaviour associated with a dynamical quantum phase transition (DQPT)~\cite{heyl2018dynamical}. Such transitions are accompanied by a non-analytic behaviour of the Loschmidt echo, defined as the overlap between the evolved state and the initial state, $\mathcal{L}(t)=\vert \bra{\psi_0} \ket{\psi(t)} \vert^2$, as a function of time. Equivalently, this critical behaviour can be captured by its normalised rate function, $r(t) = -\frac{1}{N}\log\mathcal{L}(t)$, where $N$ denotes the number of degrees of freedom of the system (e.g., number of qubits). We consider the paradigmatic example of a 1D transverse-field Ising model, described by the Hamiltonian
\begin{equation}
    H = J \sum_{i=1}^N  \sigma_z^{(i)}\sigma_z^{(i+1)}+ h \sum_{i=1}^N \sigma_x^{(i)} \, ,    \label{ising_hamiltonian}
\end{equation}
with periodic boundary conditions and $N$ denoting the number of spins in the chain.

We focus on the quantum quench where
the system is prepared in $\ket{\psi_0} = \ket{0}^{\otimes N}$ (ground state of the Ising Hamiltonian in the absence of transverse field) and then evolves under the full Hamiltonian in Eq.~\eqref{ising_hamiltonian}. 
While this system is known to exhibit a DQPT strictly in the thermodynamic limit~\cite{PhysRevLett.115.140602,PhysRevResearch.4.043161}, for finite spin chains the normalized rate function of the Loschmidt echo displays smooth peaks characteristic of quench dynamics, which serve as precursors to the DQPT as they sharpen into kinks with increasing system size.
However, sufficient noise strongly suppresses these features. 
We model the dephasing noise affecting both the $N$ spins and the ancilla qubit using the Lindblad operators $\sqrt{\gamma_z}\sigma_z$, leading to $a = N\gamma_z$.
Since these operators are all unitary up to a constant factor, we apply the simplified protocol given by the mitigation Lindbladian in Eq.~\eqref{simplified_mitigation_lindbladian}.
The normalized rate of the noiseless Loschmidt echo can then be expressed as
\begin{equation}
\begin{aligned}
    r(t) 
    &= \frac{ 2(N+1)}{N} \gamma_z + \log \Tr[(\ketbra{\psi_0}\otimes\sigma_x)W(t)] ,
\end{aligned}
\end{equation}
where $W(t)$ is the joint state of $N+1$ spins. We note that the logarithm of the expectation value from the mitigation protocol differs from the noiseless case only by a constant shift, $\tfrac{2(N+1)}{N}\gamma_z$,
thus retaining the typical signatures of a DQPT.

\begin{figure}
     \centering
     \includegraphics[width=\linewidth]{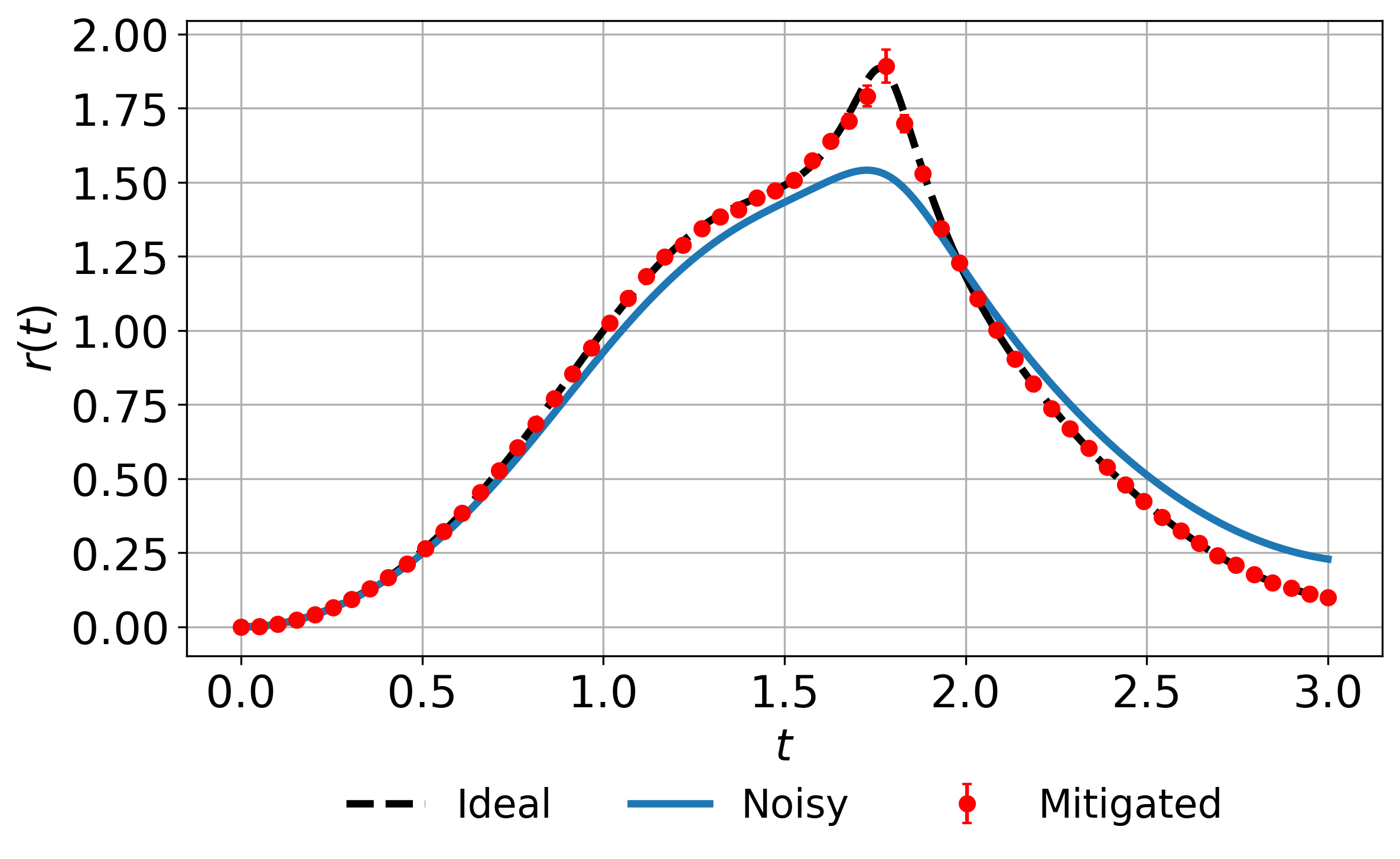}
     \caption{Normalized rate of the Loschmidt echo after quantum quench in a 1D spin Ising chain subject to the transverse field. The mitigated expectation values and their standard deviations (represented by the error bars) are obtained by averaging over $5\times10^6$ samples.}
    \label{fig:quench}
 \end{figure}

In Fig.~\ref{fig:quench} we plot the ideal, noisy, and mitigated normalized rates of Loschmidt echoes for the Ising model described above, for $N=4$, $J=0.2$, $h=1$ and $\gamma_z=0.1$. The mitigated expectation values at each time step are estimated using $5\times10^6$ samples. We observe that, while noise suppresses the 
peak, our mitigation protocol effectively recovers it. 

\subsubsection{Floquet Dynamics}
\begin{figure}
     \centering
     \includegraphics[width=\linewidth]{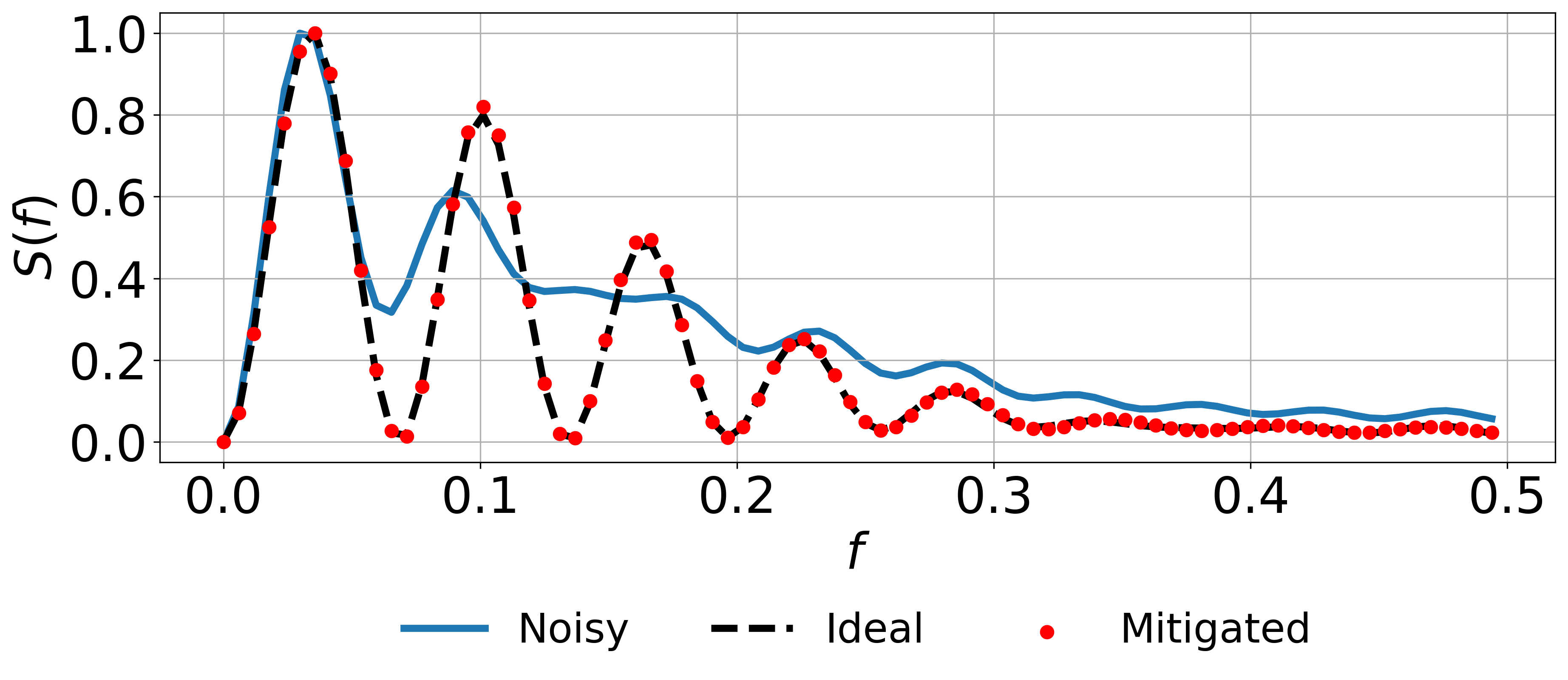}
     \caption{Normalized power spectrum of the total magnetization for a periodically-driven spin chain. The mitigated power spectrum is computed from expectation values obtained by averaging over $10^7$ samples.}
     \label{fig:floquet}
 \end{figure}

Lastly, we apply our mitigation scheme to a periodically driven (Floquet) quantum system. Specifically, we consider a spin-${1}/{2}$ chain subject to a binary drive that alternates between two non-commuting Hamiltonians,
\begin{equation}
    H_1 = J \sum_{i=1}^{N-1} \sigma^{(i)}_z \sigma^{(i+1)}_z \, ,\quad H_2 = h \sum_{i=1}^N \sigma^{(i)}_x \, ,
\end{equation}
where $N$ is the number of spins.
Each Floquet period $T$ consists of evolution under $H_1$ for a time $\delta t$, followed by evolution under $H_2$ for the remaining duration $T-\delta t$.
The system is initialized in the state $\ket{0}^{\otimes N}$, and we stroboscopically measure the expectation value of the total magnetization $\mathcal{M}=\sum_i \sigma_z^{(i)}$ at integer multiples of the driving period, $t=nT$ with $n=0,1,\dots,N_\textrm{cycles}$.
Its discrete Fourier transform $\widetilde{\mathcal{M}}(f)$, is then used to construct the (normalized) power spectrum $S_\mathcal{M}(f) \propto \vert \widetilde{\mathcal{M}}(f) \vert^2$.
In the absence of noise, the normalized power spectrum exhibits peaks at frequencies corresponding to differences of Floquet quasi-energies, reflecting coherent interference between Floquet modes, providing a direct probe of the system’s underlying quasi-energy structure.
When dephasing noise is introduced, coherences between Floquet eigenstates are progressively destroyed, leading to broadened or suppressed peaks.
To model this effect, we consider each spin to be subject to independent dephasing noise, described by jump operators $\sqrt{\gamma}\sigma^{(i)}_z$.
We then apply our mitigation scheme using the simplified Lindbladian in Eq.~\eqref{simplified_mitigation_lindbladian}.
As shown in Fig.~\ref{fig:floquet}, we compare the ideal, noisy, and mitigated power spectra of the total magnetization for a spin chain with $N=6$ spins, $\delta t =0.5$, $T=1$, $J=h=1$, $\gamma = 0.025$ and $N_\textrm{cycles}= 20$.
The mitigated expectation values of the magnetization used to compute the power spectrum are obtained by averaging over $10^7$ samples.
While dephasing washes out most of the Floquet peaks, our mitigation protocol effectively restores them, thereby recovering the correct quasi-energy structure of the driven spin system.

\section*{Discussion}
In this work, we have introduced a fully time-continuous error mitigation protocol for analogue quantum simulators that enables the recovery of noiseless expectation values of arbitrary observables. The scheme couples the system to a small number of ancillary qubits through dissipative interactions, followed by classical post-processing of the joint measurement outcomes.
In contrast to existing error-mitigation techniques that are inherently digital and therefore subject to Trotter errors and Hamiltonian-dependent artifacts~\cite{JinzhaoMitigationPaper, trajectoryQEM, ma2024limitations}, our protocol operates directly within the framework of open-system Lindbladian dynamics~\cite{Plenio_Lindblad}. As such, it is continuous in time and therefore exact for arbitrary system Hamiltonians and noise operators, avoiding both the superoperator Trotterisation errors~\cite{ma2024limitations} and the hidden dependence of parameterised noise layers on the underlying noiseless gates~\cite{JinzhaoMitigationPaper}. Our result can also be compared with a Lindbladian-based error-suppression protocol applied to encoded qubits~\cite{Kwon22reversing}, whereas our approach operates without encoding, requiring only a small number of qubits with fewer interactions, at the cost of additional sampling overhead. Moreover, we have demonstrated that the protocol remains robust under realistic ancilla noise models, which can be simply done by adjusting the post-processing factor.
Taken together, our results establish a framework for high-precision analogue quantum simulation, complement recent advances in error-mitigation strategies~\cite{filippov2023scalable, van2023probabilistic, guimaraes2024optimized}, and offer new opportunities for exploring complex quantum dynamics resilient to decoherence.

\section*{Methods}

\subsection*{Damped Schr\"odinger equation}
Here, we explicitly derive Eq.~\eqref{schrodinger_with_exp_decay} by projecting the Lindblad master equation for the joint system, Eq.~\eqref{recovery_master_equation}, onto the computational basis of the ancilla qubit.
Let $i,j\in\lbrace0,1\rbrace$, with $i\neq j$, and define $W_{ij} \equiv \bra{i}W\ket{j}$. The contribution from the system's noisy dynamics is then given by 
\begin{equation}
    \begin{split}
         & \bra{i} \mathcal{L}_\textrm{exp}(W)\ket{j}  = -i[H,W_{ij}] + \sum_k \mathcal{D}[L_k](W_{ij}) \\ & =  -i[H,W_{ij}] + \sum_k \left( L_k W_{ij} L_k^\dag  - \frac{1}{2}\lbrace L_k^\dag L_k, W_{ij} \rbrace \right) \, .
    \end{split}
\end{equation}
The mitigation Lindbladian, Eq.~\eqref{lindbladian_mitigation}, consists of three distinct terms. We compute each contribution separately, starting with
\begin{equation}
    \begin{split}
        &\bra{i}\sum_k \mathcal{D}[L_k \otimes \sigma_z] (W)\ket{j} \\
        &\quad= \sum_k \left( - L_k W_{ij} L_k^\dag  - \frac{1}{2}\lbrace L_k^\dag L_k, W_{ij} \rbrace \right)  \, ,
    \end{split}
\end{equation}
where we have used $\sigma_z^\dag \sigma_z = \mathbb{I}$. 
The remaining two contributions are given by
\begin{equation}
    \begin{split}
        \bra{i} \mathcal{D}[\sqrt{S} \otimes \mathbb{I}](W)\ket{j}=  \sqrt{S} W_{ij} \sqrt{S}- \frac{1}{2}\lbrace S, W_{ij} \rbrace  \, ,
    \end{split}
\end{equation}
\begin{equation}
    \begin{split}
        \bra{i} \mathcal{D}[\sqrt{S} \otimes \sigma_z](W)\ket{j}=  -\sqrt{S} W_{ij} \sqrt{S}- \frac{1}{2}\lbrace S, W_{ij} \rbrace   \, .
    \end{split}
\end{equation}
Combining all contributions from $\mathcal{L}_\textrm{exp}$ and $\mathcal{L}_\textrm{mit}$, we finally obtain 
\begin{equation}
\begin{split}
    & \frac{d}{dt} W_{ij} = \bra{i}\mathcal{L}_\textrm{exp}(W) + \mathcal{L}_\textrm{mit}(W)\ket{j} =   \\ &= -i[H,W_{ij}] - \lbrace \textstyle{\sum}_k L_k^\dag L_k + S, W_{ij}\rbrace \\ &=  -i[H,W_{ij}] - \lbrace a\mathbb{I},W_{ij} \rbrace = -i[H,W_{ij}] -2aW_{ij} \, ,
\end{split}
\end{equation}
where we used the definition $S = a\mathbb{I}-\sum_k L_k^\dag L_k$.
That is, $W_{ij}$ evolves according to a Schr\"odinger equation generated by $H$, with an additional damping term.

\subsection*{Noiseless expectation value}
Here, we show that the noiseless expectation value of a system observable $A$ can be recovered from the measurement on the combined system-ancilla state, evolved under the Lindblad master equation in Eq.~\eqref{recovery_master_equation}.
We previously showed that if the joint initial state is $W(0) = \rho(0) \otimes \ketbra{+}$, and we let it evolve under Eq.~\eqref{recovery_master_equation}, then the off-diagonal block of the joint density matrix reads
\begin{equation}
    W_{ij}(t) = \frac{e^{-2at}}{2}e^{-iHt}\rho(0)e^{iHt} \, .
\end{equation}
This shows that $W_{ij}(t)$ is proportional to the system's density matrix evolved under ideal unitary dynamics generated by the Hamiltonian.
To recover the noiseless expectation value of $A$, we use the fact that $W_{01}(t)=W_{10}(t)$, which follows from the symmetric initialization $W_{01}(0) = W_{10}(0)$ and that fact that both operators satisfy the same differential equation, Eq.~\eqref{diagonal_equation}.
The ideal expectation value of $A$ at time $t$ can then be written as 
\begin{equation}
\begin{split}
    \expval{A(t)}_\textrm{ideal} & = \Tr_\textrm{sys}[Ae^{-iHt}\rho(0)e^{iHt}] \\
    &= 2e^{2at}\Tr_\textrm{sys}[AW_{01}(t)] \\ & = e^{2at}\Tr_\textrm{sys}[A(W_{01}(t)+W_{10}(t))]
    \\ & 
    = e^{2at} \Tr_\textrm{sys} [A \Tr_\textrm{anc}[ (\ketbra{0}{1}+\ketbra{1}{0}) W(t)]]
    \\ & =e^{2at} \Tr[(A\otimes\sigma_x)W(t)] \, ,
\end{split}
\end{equation}
where we have used $\sigma_x = \ketbra{0}{1}+\ketbra{1}{0}$, concluding our proof.

\section*{Acknowledgments}
We acknowledge funding from the UK EPSRC through EP/Z53318X/1 and EP/W032643/1, the KIST through Open Innovation fund and the National Research Foundation of Korea grant funded by the Korean government (MSIT) (No. RS-2024-00413957). H.K. is supported by the KIAS Individual Grant No. CG085302 at Korea Institute for Advanced Study and National Research Foundation of Korea (Grants No. RS-2023-NR119931, No. RS-2024-00413957 and No. RS-2024- 00438415) funded by the Korean Government (MSIT).
\bibliography{biblio}

\end{document}


\title{Supplementary Information}
\maketitle

\section{Alternative Protocols}
The continuous-time error mitigation scheme presented in the main text can be readily adapted to suit different experimental needs. Here, we introduce several alternative protocols that demonstrate this flexibility. These include variants that: (i) retain a single qubit ancilla, but use different correlated noise dissipators; (ii) employ multiple ancilla qubits; or (iii) employ a multi-level ancilla.
All protocols share the same underlying noisy system dynamics, generated by the Lindbladian
\begin{equation}
    \mathcal{L}_\textrm{exp}(W) = -i[H\otimes \mathbb{I},W] + \sum_k \mathcal{D}[L_k \otimes \mathbb{I}](W) \, ,
\end{equation}
which acts trivially on the ancilla. 
Hence, in the following we only specify the corresponding mitigation Lindbladian $\mathcal{L}_\textrm{mit}(W)$, where $W(t)$ the joint system-ancilla density matrix.

\subsection*{Qubit ancilla}

An alternative mitigation scheme that uses a single qubit ancilla is defined by 
\begin{equation}
\mathcal{L}_\textrm{mit} = \mathcal{D}[\sqrt{2(a\mathbb{I}-\textstyle\sum_kL_k^\dag L_k)}\otimes\ketbra{0}] + \mathcal{D}[\sqrt{{2}(a\mathbb{I}-\textstyle\sum_kL_k^\dag L_k)}\otimes\ketbra{1}] +\sum_k \mathcal{D}[L_k\otimes\sigma_z] \, , 
\end{equation}
where $a$ is chosen as the largest eigenvalue of $\sum_k L_k^\dag L_k$, $ a = \lambda_\textrm{max}(\sum_k L_k^\dag L_k)$.
We initialize the joint system-ancilla state as $W(0) = \rho(0) \otimes\ketbra{+}$, and the noiseless expectation value of a system observable $A$ is recovered via
\begin{equation}    
    \langle A(t) \rangle   =
    e^{2at} \Tr[(A\otimes{\sigma_x})W(t)] \, ,
\end{equation}
resulting in the same sampling overhead found in the main text.

The protocol is also flexible when it comes to the number of ancilla qubits it employs.
In fact, although we have shown that our protocol has the desirable property of only requiring one qubit regardless of the system's dimension, certain experimental scenarios may benefit from employing multiple ancilla qubits instead.
For instance, in quantum simulators with limited connectivity, it may be difficult to implement the required correlated-noise dissipators between a single ancilla qubit and all system qubits. In such cases, our protocol can be naturally extended to accommodate multiple ancilla qubits.
As an illustrative case, let us consider a setting where the $\ell-$th system qubit is subject to local noise characterized by the Lindblad operators $\lbrace L_k^{(\ell)} \rbrace$. Furthermore, we assume that each system qubit is paired with its own dedicated ancilla qubit it can interact with (intermediate configurations are easily derived). 
In this setup, the mitigation Lindbladian becomes: 
\begin{equation}
    \mathcal{L}_{\textrm{mit}} = \sum_\ell \left( \mathcal{D}[\sqrt{S^{(\ell)}}\otimes \sigma_z^{(\ell)}] + \mathcal{D}[\sqrt{S^{(\ell)}}\otimes \mathbb{I}] + \sum_k \mathcal{D}[L_k ^{(\ell)}\otimes\sigma_z^{(\ell)}] \right) \, ,
\end{equation}
where $S^{(\ell)}=2(a^{(\ell)}\mathbb{I}-\sum_k L^{(\ell)\dag}_k L_k^{(\ell)})$, and the superscript $(\ell)$ labels each system-ancilla qubit qubit pair.
Similar to the previous protocols, we set $a^{(\ell)} =\lambda_\textrm{max}(\sum_k L_k^{(\ell)\dag} L_k^{(\ell)})$ and initialize the joint system-ancilla state as $W(0) = \rho(0) \otimes\ketbra{+}^{\otimes n}$,
where $n$ denotes the number of system qubits, and therefore that of ancilla qubits.
This symmetric initialization ensures that all $2^n$ off diagonal blocks of the joint density matrix $W$, relative to the computational basis of the ancilla qubits, coincide at all times.
As a result, the noiseless expectation value of a system observable $A$ can be recovered by jointly measuring $A$ on the system qubits, and $\sigma_x$ on each ancilla qubit
\begin{equation}
    \langle A(t)\rangle = e^{2\sum_\ell a^{(\ell)}t} \Tr[(A\otimes\sigma_x^{\otimes n})W(t)] \, ,
\end{equation}
from which it follows that the sampling overhead scales as $e^{4\sum_\ell a^{(\ell)}t}$.
A natural question is how the multi-ancilla protocol’s sampling overhead compares to that of the single-ancilla scheme.
This reduces to comparing $a=\lambda_\textrm{max}(\sum_{k\ell} L_k^{(\ell)\dag} L_k^{(\ell)})$ with $\sum_\ell a^{(\ell)} = \sum_\ell \lambda_\textrm{max}(\sum_k L_k^{(\ell)\dag} L_k^{(\ell)})$.
In general, the largest eigenvalue of a sum of positive semi-definite operators is less or equal than the sum of their individual largest eigenvalues.
Here, however, the the equality trivially holds because of the locality of the noise operators. 
Hence, we conclude that the two protocols incur in the same sampling overhead.
This changes, naturally, if the ancilla qubits are themselves noisy. In this case, the total noise to be cancelled effectively doubles, compared to the single-ancilla protocol, resulting in a quadratic increase of the sampling overhead.

\subsection*{Qutrit ancilla}
The error cancellation protocol can also be adapted to use a qutrit ancilla. 
Consider a three-level ancilla, whose Hilbert space is spanned by $\lbrace \ket{1},\ket{2},\ket{3}\rbrace$. 
A possible mitigation Lindbladian in this setting is given by
\begin{equation}
    \mathcal{L}_\textrm{mit}= \mathcal{D}[\sqrt{2(a\mathbb{I}-\textstyle\sum_iL_i^\dag L_i)}\otimes \ketbra{3}{1}]+  \mathcal{D}[\sqrt{2(a\mathbb{I}-\textstyle\sum_iL_i^\dag L_i)}\otimes \ketbra{3}{2}]+  \sum_i \mathcal{D}[L_i \otimes \tilde\sigma_z] \, ,
\label{qutrit_mitigation_lindbladian}
\end{equation}
where $\tilde\sigma_z = \ketbra{1}-\ketbra{2}$. 
The joint system-ancilla state is initialized as $W(0) = \rho (0) \otimes \ketbra{\chi}$, with $\ket{\chi} = (\ket{1}+\ket{2})/\sqrt{2}$. 
Finally, by following the same derivation steps used in the ancilla qubit case$-$now focusing on the off-diagonal blocks $\bra{i}W\ket{j}\equiv W_{ij}$, with $i\neq j$ and $i,j\in\lbrace 1,2\rbrace$, which encode the information about the noiseless system dynamics$-$we find that the noiseless expectation value of $A$ can again be expressed as
\begin{equation}
    \langle A(t)\rangle  = e^{2at} \Tr[(A\otimes \tilde\sigma_x)W(t)] \, ,
    \label{qutrit_scheme_exp_val}
\end{equation}
where $\tilde\sigma_x = \ketbra{1}{2}+\ketbra{2}{1}$. 
Alternative mitigation Lindbladians to Eq.~\eqref{qutrit_mitigation_lindbladian} can easily be constructed by inspection.
For instance,
the operator $\ketbra{3}{1}$ may be replaced with any $\ketbra{\psi}{1}$ such that $\bra{\psi}\ket{1} = 0$ and/or  $\bra{\psi}\ket{2} = 0$; similarly  $\ketbra{3}{2}$ may be replaced with any operator $\ketbra{\phi}{1}$ satisfying $\bra{\phi}\ket{2} = 0$ and/or  $\bra{\phi}\ket{1} = 0$.

Similarly to the qubit ancilla protocol discussed in the main text, this qutrit ancilla protocol also exhibits robustness under certain types of ancilla noise.
For simplicity, we model the ancilla noise using a single Lindblad noise operator $M$. As a result, the master equation governing the evolution of the joint system-ancilla state becomes
\begin{equation}
\label{full_dynamics}
    \frac{d}{dt}W(t) = \mathcal{L}_\textrm{exp}(W) + \mathcal{L}_\textrm{mit}(W) + \mathcal{L}_\textrm{anc}(W) \, ,
\end{equation}
where 
\begin{equation}
    \mathcal{L}_\textrm{anc}(W)=\mathcal{D}[\mathbb{I}\otimes M](W) \, .
    \label{ancilla_noise_lindbladian}
\end{equation}
Consequently, the differential equation governing the evolution of $W_{ij}$ now includes the additional term $\bra{i}\mathcal{D}[\mathbb{I}\otimes M](W)\ket{j}$.
The form of the operator $M$ largely depends on the specific physical implementation of the three-level systems and, in particular, on the energy level arrangement and transition dynamics between the three states. 
In general, we can typical ancilla noise processes to include pure dephasing terms of the form $\ketbra{i}$, as well as excitation loss or heating terms such as $\ketbra{i}{j}$, with $i\neq j $. 
By substituting these noise operators into Eq.~\eqref{ancilla_noise_lindbladian} and computing their contributions excplicitly, we find that only two types of behaviour arise.
In some cases$-$specifically for the operators $\lbrace \ketbra{1}{3},\ketbra{2}{3}$, and$\ketbra{3} \rbrace-$ the contribution vanishes.
In all other cases, i.e., for the remaining six operators, the noise introduce a non-zero contribution that reads $-W_{ij}/2$. 
As shown in the qubit-ancilla case, these noise contributions do not affect the correlated-noise dissipators within $\mathcal{L}_\textrm{mit}$. Instead, they simply modify the exponential pre-factor appearing in Eq.~\eqref{qutrit_scheme_exp_val}.
In other words, these ancilla noise effects can be accounted for through a classical post-processing, namely an appropriate redefinition of $e^{2at}\mapsto e^{2\tilde{a}t}$, which only affects the sampling overhead of the protocol without altering its underlying structure.

\section{Robustness against correlated noise}
The mitigation dynamics described in Eq.~\eqref{full_dynamics} can also handle certain forms of correlated noise between the system and the ancilla. For instance, let us consider a Lindbladian noise term $\mathcal{D}[P \otimes \sigma_\pm]$, where $P$ is a generic Pauli string acting on the system qubits. A straightforward calculation shows that $\bra{i}\mathcal{D}[ P \otimes \sigma_\pm](W)\ket{j}= - W_{ij}/2$, implying that the effect of this noise can be cancelled out via classical post-processing, as shown in the main text.
If we instead consider $\mathcal{D}[P \otimes \sigma_z]$ as model of correlated noise, we easily find $\bra{i}\mathcal{D}[ P \otimes \sigma_z](W)\ket{j}= -PW_{ij}P - W_{ij}$ which, contrary to the previous case, is not proportional to $W_{ij}$, implying that mitigation via our scheme is not possible by simple post-processing in this scenario. Nevertheless, by introducing additional system noise via $\mathcal{D}[P\otimes \mathbb{I}]$, which contributes with $\bra{i}\mathcal{D}[ P \otimes \mathbb{I}](W)\ket{j}= PW_{ij}P - W_{ij}$ to the master equation governing $W_{ij}$, the combined effect is now proportional to $W_{ij}$ and reads $-2W_{ij}$.
This shows that even this form of correlated noise can be mitigated with an appropriate modification of the scheme.
A comprehensive analysis of the protocol’s robustness to ancilla noise, beyond the specific examples considered here, is left for future work.